\input epsf
\documentstyle[preprint,aps,pre]{revtex}     
\tighten
\begin{document}
\draft


\title{Near-Surface Long-Range Order at the Ordinary Transition\vspace*{0.6cm}}
\author{Uwe Ritschel\footnote{e-mail: uwe@theo-phys.uni-essen.de}
 and Peter Czerner\footnote{e-mail: peterc@theo-phys.uni-essen.de}\vspace*{0.6cm}}
\address{Fachbereich Physik, Universit\"at GH Essen, 45117 Essen
(F\ R\ Germany)\vspace*{1.8cm}}
\date{submitted to Phys. Rev. Lett.\vspace*{1.6cm}}

\maketitle
\narrowtext
\begin{abstract}
We study the spatial dependence of the order parameter near
surfaces belonging to the universality class of the ordinary
transition.
Special attention is paid to the influence of a small
surface magnetic field
$h_1$ at and above the bulk critical temperature. 
A detailed scaling analysis (which is
confirmed by a perturbative calculation)
reveals that $h_1$ may give rise to an anomalous short-distance
behavior of the order parameter. Close to the surface
the magnetization {\it increases}
with a power law $m\sim z^{\,\kappa}$ with $\kappa\equiv
1-\eta_{\perp}^{ord}\simeq 0.23$ for the three-dimensional Ising model. 
These results are closely related to
experimental findings where exponents
of the ordinary transition were observed in Fe$_3\,$Al,
while superstructure reflections revealed the existence of
long-range order near the surface [X. Mail\"ander et al.,
Phys. Rev. Lett. {\bf 64}, 2527 (1990)].
\end{abstract}
\pacs{PACS: 75.40.Cx,75.30.Pd,78.70.Ck,68.35.Rh}
A prototypical 
system to study critical phenomena in restricted geometries
is the semi-infinite 
Ising model, terminated by a plane surface and extending infinitely 
in the direction perpendicular to the surface ($z$-direction) \cite{binder}.
Spins located in the surface may experience interactions different
from those in the bulk, for example due to missing neighbors
at a free surface or due to a strong coupling to an adjacent medium.
In the framework of continuum field theory such as 
the $\phi^4$ model the surface influence is
taken into account by additional fields like the surface magnetic field
$h_1$ and the local temperature perturbation $c_0$ at $z=0$. The latter can be
related to the surface enhancement of the spin-spin coupling in
lattice models \cite{diehl}.

At the bulk critical temperature
$T_c^b$ the tendency to order near the surface
can be reduced ($c_0>0$) or increased ($c_0<0$),
or, as a third possibility,
the surface can be critical as well. As a result, each
bulk universality
class in general divides into several distinct
surface universality classes, called
ordinary ($c_0\rightarrow \infty$), extraordinary ($c_0\rightarrow -\infty$),
and special transition ($c_0=c_{sp}^*$).

Close to the surface, within the range of the bulk correlation length $\xi
\sim |\tau|^{-\nu}$,
the singular behavior of thermodynamic quantities is markedly changed
compared to the bulk. For 
$z\ll \xi$ the magnetization behaves as
$\sim |\tau|^{\beta_1}$ when $\tau=(T-T_c^b)/T_c^b\to 0$
from below, with $\beta_1$
assuming characteristic values for special and ordinary transition, which
are in general different from the bulk exponent $\beta$. (At the extraordinary
transition the surface is already ordered at $T_c^b$.) Further, the
correlation functions
near the surface are characteristically modified.
The correlation function for points
within a plane parallel 
to the surface 
is given by $C(\Delta{\bf r}_{\parallel}) \sim 
|\Delta{\bf r}_{\parallel}
|^{-(d-2+\eta_{\parallel})}$, where the anomalous
dimension $\eta_{\parallel}$ is related to $\beta_1$ by
$\beta_1=(\nu/2)(d-2+\eta_{\parallel})$ \cite{diehl}. Correlations
in the z-direction (and all other directions except the parallel one)
are governed by $C(z,z')\sim |z-z'|^{-(d-2+\eta_{\perp})}$.

Some of the theoretical predictions \cite{diehl,ord}
were found in excellent agreement with experiments carried
out by Mail\"ander et al. \cite{mail,dosch}. In these experiments
Fe$_3$\,Al was studied close to the $D0_3$-$B_2$ transition
by scattering of evanescent waves generated
by total reflection of x-rays at a
$[1\underline{1}0]$ surface. The system was expected to
belong to the universality class of the
ordinary transition, and indeed
the exponents
measured were in remarkable
agreement with theoretical predictions \cite{mail}. 
A somewhat disturbing feature was
that superstructure reflections revealed the existence of unexpected
long-range order (LRO) near the surface reminiscent to the situation
at the extraordinary transition. In the sequel it was demonstrated
by Schmid \cite{schmid} that in a similar situation (at the
$A_2$-$B_2$ transition in Fe$_3$Al)
an effective ordering field $h_1$ at the surface can arise when
the stoechiometry of the alloy is not ideal. Assuming
that an $h_1$ is also present at the $DO_3$-$B_2$ transition, the
observed LRO can be explained, leaving
unanswered the question, however, why
exponents of the ordinary transition were measured despite the LRO.
In the following we show that
a {\it small} $h_1$ may generate an universal
power-law growth of the order parameter near the surface and, 
as a result, a LRO considerably (and, in fact, infinitely)
larger than expected from mean-field (MF) approximations,
while the correlation
function is still governed by the exponents of the ordinary transition.

Most of the theoretical studies 
concerning inhomogenous systems
concentrated on the behavior at the
fixed points $c_0=\pm\infty$ and $c_0=c_{sp}^*$, respectively. 
At $T_c^b$ for both the ordinary 
and the special transition (for $h_1=0$)
the order-parameter profiles are zero for all $z\ge 0$. At the 
extraordinary transition the surface is ordered and the order decays 
as $z^{-\beta/\nu}$ with increasing distance
from the surface \cite{foot}, 
where in the Ising-case $\beta/\nu \simeq 0.52$ \cite{ruge}.
Concerning the effects of $h_1$
it was assumed for a long time
\cite{bray} and recently also shown by rigorous
arguments \cite{dibu}
that a strong $h_1$ at the ordinary transition 
(the so-called normal transition)
is equivalent to the extraordinary transition.
The special transition was studied by Brezin and Leibler
\cite{brezin} and by Ciach and Diehl \cite{ciach}. It was found that
at the fixed point the scaling field
$h_1$ gives rise to a length scale $ l^{sp}\sim
h_1^{-\nu/\beta_1^{sp}}$.
For $z\gg l^{sp}$ one finds that $m\sim z^{-\beta/\nu}$  
as at the extraordinary transition. In the opposite limit,
$z\ll l^{sp}$, the magnetization behaves as $m\sim 
m_1\,z^{(\beta_1^{sp}-\beta)/\nu}$.
Since $\beta_1^{sp}\le \beta$, the order still decays, governed
by a somewhat smaller exponent
compared to large distances. For the Ising model
$(\beta_1^{sp}-\beta)/\nu \simeq -0.15$ \cite{ruge}.

What can we expect if a {\it small}
$h_1$ is applied in the presence of a {\it large}
$c_0$, i.e. close to the fixed-point of the ordinary transition.
In this situation the
parameter $c_0$ is a so-called dangerous
irrelevant variable \cite{diehl,gompp}, 
comparable to the $\phi^4$ coupling constant $g$
at and above the upper
critical dimension $d^{*}=4$, and in general must not be naively set to
its fixed point value $c_0=\infty$.   
Setting the bulk magnetic field $h=0$, the remaining linear
scaling fields at the ordinary transition
are $\tau$ and ${\sf h}_1\equiv h_1/c_0$
\cite{diehl,gompp,foot2}. Hence,
the behavior of the magnetization under
rescaling of distances should be described by
\begin{equation}\label{scal}
m(z,\tau,{\sf h}_1)\sim b^{-\beta/\nu}\,m(zb^{-1},\,\tau b^{1/\nu},\,
{\sf h}_1\,b^{x_1}),
\end{equation}
where the scaling dimension of ${\sf h}_1$ is given by
$x_1=(d-\eta_{\parallel}^{ord})/2$ \cite{diehl}.

Let us first discuss the profile
for ${\sf h}_1=0$. As mentioned above, for $\tau>0$ we have
$m=0$ everywhere. For $\tau<0$, on the other hand, the magnetization approaches
its bulk value $m_b\sim |\tau|^{\beta/\nu}$ for $z\to\infty$.
Close to the surface ($z\ll \xi$) the magnetization
increases with a power law \cite{gompper}. To see this from (\ref{scal}), we
set ${\sf h}_1=0$ and fix the arbitrary rescaling parameter $b$
by setting it $\sim z$.
Then the magnetization takes the scaling form
\begin{equation}\label{tau}
m(z,\tau)\sim z^{-\beta/\nu}\,{\cal M}_{\tau}(z/\xi)\>.
\end{equation}
Since we expect that $m(z\to 0)\sim m_1$ and know that $m_1\sim
|\tau|^{\beta^{ord}_1}$, we conclude
for the short-distance form of the scaling function 
${\cal M}_{\tau}\sim \zeta^{\beta^{ord}_1/\nu}$
and, in turn, the behavior of $m$ is given by
$m(z,\tau)\sim |\tau|^{\beta_1^{ord}}\,
z^{({\beta^{ord}_1-\beta})/\nu}$ \cite{gompper}.

We now turn to the case $\tau=0$ and ${\sf h}_1\neq 0$. This is
the situation we are actually interested in and which is important
for understanding the experimental results of Ref. \cite{mail}.
In this case the scaling form
derived from (\ref{scal}) is
\begin{equation}\label{h1}
m(z,{\sf h}_1)\sim z^{-\beta/\nu}\,{\cal M}_{{\sf h}_1}(z{\sf h}_1^{1/x_1})\>.
\end{equation}
First of all we notice from (\ref{h1}) that the scaling field
${\sf h}_1$ gives rise to a length scale
$l^{ord}\sim {\sf h}_1^{-1/x_1}$ quite
comparable to the situation near the special transition discussed above.
In order to find the short-distance
behavior of ${\cal M}_{{\sf h}_1}(\zeta)$
we have to recall that the surface is paramagnetic at the
ordinary transition.
Thus $m_1$ 
will respond linearly to ${\sf h}_1$ \cite{foot3}. Arguing again that
$m(z\to 0)\sim m_1$, we now find
that ${\cal M}_{{\sf h}_1}\sim \zeta^{x_1}$ for 
$\zeta\ll 1$, and, in turn, with the scaling relation 
$\eta_{\perp}=(\eta_{\parallel}+\eta)/2$ \cite{diehl},
the short-distance behavior is given by
\begin{equation}\label{sdb}
m(z,{\sf h}_1)\sim {\sf h}_1\,z^{\kappa} \qquad \mbox{with}\qquad
\kappa\equiv 1-\eta^{ord}_{\perp}\>.
\end{equation}
In the opposite limit, $z\gg l^{ord}$, the magnetization approaches
the bulk equilibrium value zero as $\sim z^{-\beta/\nu}$. 

Eq.\,(\ref{sdb}) is the central result of this Letter. It states that
the magnetization even at (or slightly above) $T_c^b$ in the presence
of a surface field $h_1$ shows a power-law increase reminiscent
of the situation below $T_c^b$. The short-distance exponent $\kappa$
defined in (\ref{sdb}) is zero in MF theory. Below $d^*$,
however, as for the Ising system in $d=3$, it is nonzero
and positive. Taking the literature values for the surface exponents from
Ref. \cite{diehl}, one obtains $\kappa\simeq 0.23$, which implies a
rapid growth of LR0 with increasing $z$.

The {\it spatial} variation of the
magnetization discussed above
strongly resembles the {\it time} dependence of the
magnetization in relaxational processes at the critical point.
If a system with nonconserved order parameter (model A) is quenched from a
high-temperature initial state to the critical point, with a small
initial magnetization $m^{(i)}$, the order parameter behaves as $m \sim
m^{(i)}\,t^{\theta}$ \cite{jans}, where the short-time exponent $\theta$
is governed
by the difference between the scaling dimensions of initial
and equilibrium magnetization \cite{own}. Like the exponent $\kappa$ 
in (\ref{sdb}), the exponent 
$\theta$ vanishes in MF theory, but becomes positive
below $d^*$.

There is also a heuristic argument for the growth of LRO
near the surface. As said above, $h_1$
generates a surface magnetization $m_1\sim h_1$. 
Regions (on macroscopic scales) close to the surface
will respond to this magnetization by ordering as well. How strong this
influence is depends on two factors. First, it is proportional to
the correlated area in a plane parallel to the surface in a distance $z$.
While in the surface correlations are suppressed,
close to the surface the correlation length 
in directions parallel to the surface,
$\xi_{\parallel}$, grows as $\sim z$ \cite{ciach}. 
Second, for small $h_1$ (and thus small surface magnetization) 
it depends linearly on the probability that a given spin orientation
``survives'' in a distance $z$ from the surface.
The latter is governed by the {\it perpendicular}
correlation function $C(z)\sim z^{-(d-2+\eta_{\perp}^{ord})}$.
Taking the factors together, we obtain
\begin{equation}\label{heuristic}
m(z)\sim h_1\,C(z)\,\xi_{\parallel}^{d-1}=h_1\,z^{1-\eta_{\perp}^{ord}}.
\end{equation}
Qualitatively speaking, 
the surface when carrying a small $m_1$ induces a much larger
magnetization in the adjacent layers, which are much more
susceptible and respond with a
magnetization $m(z)\gg m_1$. This effect is not observed
on the MF level since there the increase of the correlated surface
area is {\it exactly}
compensated by the decay of the perpendicular correlations. 

In order to corroborate our scaling analysis and the heuristic
arguments from above, we carried out
a one-loop calculation for the $\phi^4$-model
employing the $\epsilon$ expansion. Expanded in powers of the coupling
constant, the magnetization can be written in the form
$m=m^{(0)}+g\,m^{(1)}+{\cal O}(g^2)$, where $m^{(0)}$ is the well known MF solution
\cite{bray,lubrub} and $m^{(1)}$ is the 1-loop term. The latter
was calculated exactly for arbitrary $c_0$ and $h_1$ 
in Refs.\,\cite{brezin,ciach}. However, the renormalization of
uv-divergences and the following improvement with the help
of the renormalization group where done at (or in the vicinity of)
the special transition in these references. As a consequence, the
anomalous short-distance behavior at the ordinary transition was 
missed.

The MF solution that satisfies the boundary condition
$\left. \partial_z\,m-c\,m\right|_{z=0}=h_1$
at the surface is given by
\begin{mathletters}\label{mfsolu}
\begin{equation}\label{mffunct}
m^{(0)}(z)=\sqrt{\frac{12}{g}}\,\frac{1}{\tilde z}
\end{equation}
with
\begin{equation}\label{length}
\tilde z\equiv z+z_+\quad \mbox{and}\quad
z_+^{-1}=\frac{\left(c_0^2+4 h_1\,\sqrt{g/12}
\right)^{1/2}-c_0}{2}\>,
\end{equation}
\end{mathletters}
which holds for general $c_0$ and $h_1$. Close to the ordinary transition
(large $c_0$) the mean field length scale becomes $z_+\simeq 
l^{ord}=(12/g)^{1/2}\,c_0/h_1$.
As expected
from (\ref{sdb}), there is no anomalous short-distance behavior
on the MF level. The profile has its maximum value at $z=0$, and
for $z\gg l^{ord}$ the
profile decays as $\sim z^{-\beta/\nu}$ with the MF value
$\beta/\nu=1$.

The one-loop term $m^{(1)}$ is given by \cite{ciach,smock}
\begin{equation}\label{oneloop}
m^{(1)}(z)=-\frac{1}{2}\int_0^{\infty}\,dz'\;C(0;z,z')\,m^{(0)}(z')\,
\int_pC(p;z',z')\>,
\end{equation}
where $m^{(0)}(z)$ is the zero-loop (MF) profile (\ref{mffunct}) and
$\int_p\equiv (2\pi)^{1-d}\int d^{d-1}p$.
The propagator $C(p;z,z')$ is Fourier-transformed with respect
to the spatial coordinates parallel to surface. It can be calculated
exactly \cite{brezin,ciach}, and the somewhat lenghty
results will be omitted here. 
The integrations in (\ref{oneloop}) necessary
to obtain the full scaling function
${\cal M}_{{\sf h}_1}$ are
complicated and can only be carried out numerically.
However, it is straightforward to extract the divergent terms, poles
$\sim 1/\epsilon$ in dimensional regularization, and the
short-distance singularities $\sim \log z$, which, when
exponentiated, give rise to power laws modified compared to the
MF theory.
Collecting these terms, $m^{(1)}$ is given by (\ref{oneloop}) with 
\begin{equation}\label{idiv}
\int_pC(p;z,z) =
\frac{K_{d-1}}{2}\,\tilde z^{-2+\epsilon}\,
\int_1^{\infty}dy\,y^{1-\epsilon}
\left[
e^{-2y}\left(e^{2y l^{ord}/\tilde z}-1\right)
\left(1+\frac{3}{y}+\frac{3}{y^2}\right)^2
-\frac{3}{y^2}\right]
+\mbox{finite}
\end{equation}
where $K_d\equiv 2/\left((4\pi)^{d/2}\Gamma(d/2)\right)$ and
``finite'' stands for terms which are
finite for $\epsilon\to 0$ and $z\to 0$. Terms of ${\cal O}(1/c_0)$
are also omitted in (\ref{idiv}).
The zero-momentum
propagator $C(0,z,z')$ appearing in (\ref{oneloop})
(for $c_0\to \infty$) takes the simple form
\begin{equation}\label{zeromom}
C(0,z,z')=\frac{1}{5}\,\frac{1}{\tilde z_{<}^2\,\tilde z_{>}^2}
\left(\tilde z_{<}^5-\tilde z_+^5\right)\>,
\end{equation}
where $<(>)$ denotes the smaller (larger) of $\tilde z$ and $\tilde z'$.

Further analysis shows that the uv-divergences can be absorbed
in the standard fashion
by renormalization of the coupling constant $K_d g_0=
u\,(1+3u/2+{\cal O}(u^2))$ and of the scaling field ${\sf h}_{1,0}
={\sf h}_1(1-u/4\epsilon+{\cal O}(u^2))$ \cite{diehl}.
After this the
coupling constant is set to its fixed point value $u^*=2\epsilon/3$.
Eventually, after exponentiation
of logarithms we find the asymptotic power laws
\begin{equation}\label{asympt}
m(z,{\sf h}_1)\sim \left\{
\begin{array}{lll}
z^{-1+\epsilon/2} & \mbox{for} &z\gg l^{ord}\\
{\sf h}_1\,z^{\epsilon/6} & \mbox{for} &z\ll l^{ord}\>.
\end{array}\right.
\end{equation}
As expected the decay of the
profile for $z\gg l^{ord}$
is governed by the one-loop result $\beta/\nu = 1-\epsilon/2$.
The short-distance behavior is consistent with our
scaling analysis; in first-order $\epsilon$ expansion $\kappa=
1-\eta_{\perp}^{ord}=\epsilon/6$ \cite{diehl}.

\def\epsfsize#1#2{0.9#1}
\hspace*{1cm}\epsfbox{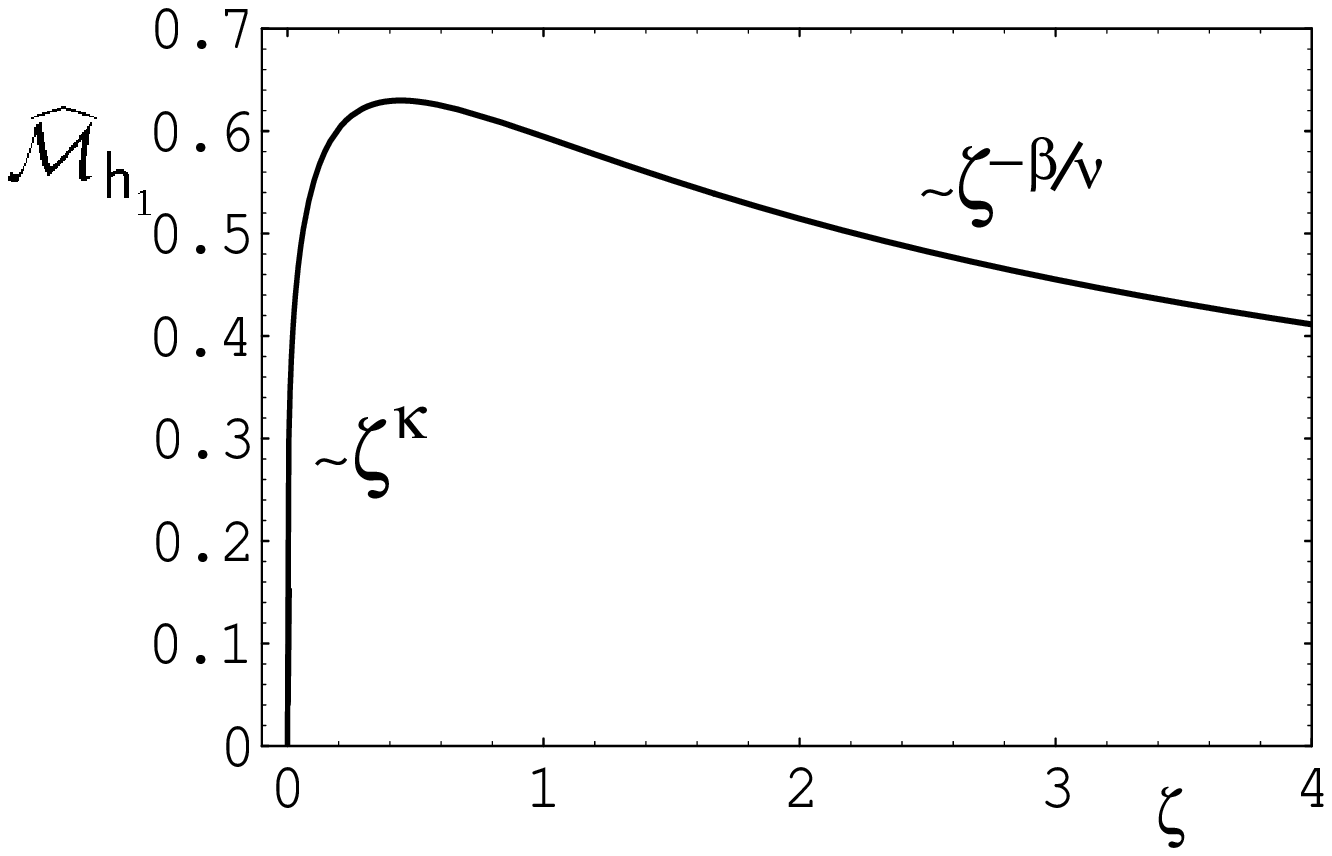}\\[0mm]
{\small {\bf Fig.\,1:} Qualitative shape of the scaling
function $\widehat {\cal M}_{{\sf h}_1}(\zeta)=
\zeta^{-\beta/\nu}{\cal M}_{{\sf h}_1}(\zeta)$
of the magnetization. More details are described in the text.}\\[0.1cm]

A more detailed account concerning the behavior of the magnetization
in between the asymptotic regimes of Eq.\,(\ref{asympt}) will be given
elsewhere \cite{czeri}. A qualitative preview on the form of the
scaling function $\widehat {\cal M}_{{\sf h}_1}(\zeta)\equiv
\zeta^{-\beta/\nu}{\cal M}_{{\sf h}_1}(\zeta)$ (see Eq.\,(\ref{h1}))
is shown in Fig.\,1, where the asymptotic power laws are quantitatively
correct but the crossover is described by a simple substitute
for the scaling function. Regarding the crossover between ordinary
(${\sf h}_1=0$) and the extraordinary (or
normal) transition (${\sf h}_1=\infty$) the
following scenario should hold. While at the ordinary transition $m(z)$
vanishes everywhere, for ${\sf h}_1\neq 0$ the magnetization
increases as $\sim z^{\kappa}$ up to $z\simeq
l^{ord}\sim {\sf h}_1^{-1/x_1}$ ($1/x_1\simeq 1.33$ for the Ising model)
and thereafter crosses over to the long-distance form given
in (\ref{asympt}). When ${\sf h}_1$ becomes larger, the short-distance
increase is steeper and $l^{ord}$ shrinks. For ${\sf h}_1\to
\infty$ we have $l^{ord}\to 0$, and one finds $m\sim z^{-\beta/\nu}$ for
all (macroscopic) distances, the
result at the extraordinary transition.
Largely analogous results---monotonous behavior at the fixed points
and profiles with one extremum in the crossover regime---were found
by Mikheev and Fisher for the energy density of the two-dimensional
Ising model \cite{mifi}.

In conclusion, we studied the effects of a small surface magnetic
field at the ordinary transition at or above $T_c^b$.
We found order-parameter exhibits
anomalous short-distance behavior in form of a power-law
increase $m\sim h_1 z^{1-\eta_{\perp}^{ord}}$ near the surface
implying a much more pronounced long-range order than could be expected
from mean-field theory. For $z$ smaller than $l^{ord}$ and $\xi$, the
magnetization increases, while the correlation function (and
thus the structure function) is governed by the asymptotic form
of the ordinary transition up to correction of ${\cal O}(1/c_0)$ \cite{eisen}.
Assuming that there exists a small ordering field $h_1$ in the
system studied by Mail\"ander et al. \cite{mail}, our scenario
gives a plausible explanation for the experimental findings.
 
{\small {\it Acknowledgements}:
We should like to thank H. W. Diehl,
E. Eisenriegler, and R. Leidl for useful discussions and
hints to the literature. Especially we thank H. W. Diehl for
leaving us the manuscript Ref. \cite{ciach} prior to publication.
This work was supported in part by the Deutsche Forschungsgemeinschaft
through Sonderforschungsbereich 237.}


\begin{thebibliography} {99}
\bibitem{foot} From the power-law decay $m\sim z^{-\beta/\nu}$ at the
extraordinary transition it appears that the magnetization diverges
when going towards the surface. One has to bear in mind however that the
power law is only valid for distances much larger than microscopic scales.
Upon approaching the surface, the magnetization would depart from the power
law and assume a finite value at the surface.
\bibitem{foot2} The correct scaling field 
at the ordinary transition is actually $h_1/c_0^y$, where
the exponent $y=1$ in MF theory but $\neq 1$ in general. It is discussed
in detail in Ref.\,\cite{diehl} that in the framework of the loop
expansion one does not capture the deviation from the MF value
in this exponent, while e.g. the $z$-dependence 
of expectation values is reproduced correctly.
For details we refer to Refs. \cite{diehl,gompp}.
\bibitem{foot3} The relation $m_1\sim {\sf h}_1$ seems to be inconsistent
with the definition of the surface
exponent $\delta_{11}$ via $m_1\sim h_1^{1/\delta_{11}}$ and
the result $\delta^{ord}_{11}\simeq 0.6$ for the Ising model \cite{diehl}.
One has to bear in mind, however, that $\delta^{ord}_{11}$
governs the leading
singular behavior of $m_1$ which is weaker here than the
linear dependence on $h_1$ \cite{bray}.
\end{thebibliography}
\end{document}